\documentclass[12pt,english]{article}
\usepackage[T1]{fontenc}
\usepackage[latin9]{inputenc}
\usepackage[a4paper]{geometry}
\usepackage{amsmath}
\usepackage{amssymb}
\usepackage{esint}

\makeatletter
\numberwithin{equation}{section}
\newcommand{\lyxaddress}[1]{
\par {\raggedright #1
\vspace{1.4em}
\noindent\par}
}

\makeatother

\usepackage{babel}
\begin{document}

\title{Explicit boundary form factors: the scaling Lee-Yang model}

\author{L. Hollo$^{1}$, Z. B. Laczko$^{2}$ and Z. Bajnok$^{1}$}

\maketitle

\lyxaddress{\begin{center}
$^{1}$MTA Lend\"ulet Holographic QFT Group, Wigner Research Centre
for Physics\\
H-1525 Budapest 114, P.O.B. 49, Hungary\\
~\\
$^{2}$ Roland Eötvös University, \\
 1117 Budapest, Pázmány Péter sétány 1/A, Hungary\\

\par\end{center}}
\begin{abstract}
We provide explicit expressions for boundary form factors in the boundary
scaling Lee-Yang model for operators with the mildest ultraviolet
behavior for all integrable boundary conditions. The form factors
of the boundary stress tensor take a determinant form, while the form
factors of the boundary primary field contain additional explicit
polynomials. 
\end{abstract}

\section{Introduction}

A complete solution of a $1+1$ dimensional integrable QFT means the
construction of all of its Wightman functions and the procedure to
reach this goal is called the bootstrap programme. 

The first step is the so-called S-matrix bootstrap which determines
the multiparticle scattering matrix. This S-matrix connects the asymptotic
in and out states, and factorizes into pairwise elastic scatterings
satisfying unitarity, crossing symmetry and the Yang-Baxter equations.
Maximal analyticity is also required: poles of the S-matrix have to
be located on the imaginary axis on the rapidity plane and have to
be explained by bound states or some Coleman-Thun diagrams (for detailed
review see \cite{Dorey:1996gd,Bajnok:2011sa,Bajnok:2013sa}). Assuming
that the only particle of the model appears also as a bound state
the S-matrix bootstrap programme results in the S-matrix of the Lee-Yang
model.

Integrable impurities with nontrivial bulk S-matrices can be either
purely transmittive defects or purely reflective boundaries \cite{Delfino:1994nr,CastroAlvaredo:2002fc}.
In these cases the S-matrix bootstrap has to be complemented by the
determination of the defect transmission or boundary reflection matrices
via the T-matrix or R-matrix bootstrap. Requiring unitarity, crossing
unitarity and the defect/boundary Yang-Baxter equations together with
maximal analiticity determines these matrices up to CDD factors which
can be fixed by some physical input \cite{Ghoshal:1993tm,Bajnok:2004jd}.

The final step in the bootstrap programme is the form factor bootstrap.
Form factors are the matrix elements of local operators between asymptotic
states. An axiomatic formulation of form factors was initiated in
\cite{Smirnov:1992vz} in the bulk and was then extended to the boundary
\cite{Bajnok:2006ze} and defect \cite{Bajnok:2009hp} cases. In the
bootstrap framework the determination of the form factors consists
of finding all solutions of the form factor axioms. These axioms are
functional relations which also connect form factors with different
particle numbers. 

In this axiomatic approach a family of form factor solutions, called
tower, corresponds to a local field. As the axioms do not contain
any information about these fields the identification is not obvious.
However, first for the Ising model \cite{Cardy:1990pc} and later
for the scaling Lee-Yang model \cite{Christe:1990zt,Koubek:1994gk,Koubek:1994zp}
the space of the form factor solutions was shown to be isomorphic
to the space of local operators (see also \cite{Smirnov:1995jp})
and later this isomorphism has been shown level by level \cite{Delfino:2007bt}.
This counting argument was then extended to the boundary Lee-Yang
model in \cite{Szots:2007jq} and also to the boundary Sinh-Gordon
model with Dirichlet boundary condition at the self dual-point \cite{CastroAlvaredo:2006sh}.

The importance of explicit form factor solutions lies in the fact
that they can be used to build up correlation functions by their spectral
representations (for review see \cite{Smirnov:1992vz}). However,
the explicit solution for towers of form factors is not easy and so
far this goal was achieved only for a few models such as for the scaling
Lee-Yang model \cite{Zamolodchikov:1990bk} and its defect version
\cite{Bajnok:2013eaa}, for the Sinh-Gordon model \cite{Fring:1992pt,Koubek:1993ke,Pillin:1997bx}
and for some homogeneous Sine-Gordon models \cite{CastroAlvaredo:2000nk,CastroAlvaredo:2000nr}.
In \cite{CastroAlvaredo:2006sh} a closed formula was conjectured
for all $n$-particle form factors for some operators in the boundary
Sinh-Gordon model with Dirichlet boundary condition at special coupling.
It is worth mentioning that in \cite{CastroAlvaredo:2007pe} the boundary
one-particle minimal form factors were calculated for the $A_{n}$
affine Toda field theories and in case of the $A_{2}$ theory the
solutions of the axioms were given up to four particles for some operators.
In this paper, for the first time, we give explicit\emph{ boundary}
form factor solutions for the lowest lying fields for all possible
integrable boundary conditions in the scaling Lee-Yang model.

The paper is organized as follows: In Section \ref{sec:Boundary-form-factors:}
we briefly review the scaling Lee-Yang model, the boundary form factor
axioms and introduce a useful Ansatz for the form factors. In Section
\ref{sec:Explicit-Boundary-Form} we give the explicit solutions in
three steps: first we give the form factor tower for the energy-momentum
tensor in case of the identity boundary in a determinant-form. Then
by the fusion method we extend this result for the other integrable
boundary conditions. Finally, we derive the form factor tower for the
remaining boundary primary field. In Section \ref{sec:Conclusion}
we briefly summarize our results and conclude. The details of the
proof for the identity boundary condition are relegated to Appendix
\ref{sec:Proof-of-the}.

\section{Boundary form factors: axioms and parametrization\label{sec:Boundary-form-factors:}}

In this section, following \cite{Bajnok:2006ze}, we recall the boundary
form factor axioms and a parametrization which fulfills them. We analyze
the scaling Lee-Yang theory, which is the simplest integrable quantum
field theory containing one particle type with mass $m$. The multiparticle
scattering matrix factorizes into pairwise scatterings, which depends
on the difference of the rapidities of the particles $S(\theta_{1}-\theta_{2})$
and takes the following simple form: 
\begin{equation}
S(\theta)=\frac{\sinh\theta+i\sin\frac{\pi}{3}}{\sinh\theta-i\sin\frac{\pi}{3}}\equiv-\left(2\right)_{\theta}\left(4\right)_{\theta}\quad;\quad\left(x\right)_{\theta}=\frac{\sinh\left(\frac{\theta}{2}+\frac{i\pi x}{12}\right)}{\sinh\left(\frac{\theta}{2}-\frac{i\pi x}{12}\right)}
\end{equation}
where $p=m\sinh\theta$. The pole of the scattering matrix at $\theta=\frac{2i\pi}{3}$:
\begin{equation}
S(\theta)=i\frac{\Gamma^{2}}{\theta-\frac{2i\pi}{3}}+\mbox{reg.}\quad;\qquad\Gamma=i\sqrt{2\sqrt{3}}
\end{equation}
signals a bound-state, while the fusion relation 
\begin{equation}
S(\theta)=S\left(\theta-i\frac{\pi}{3}\right)S\left(\theta+i\frac{\pi}{3}\right)
\end{equation}
ensures that the bound-state is the particle itself. 

In the presence of integrable boundaries the scattering matrix has
to be supplemented by the one particle reflection factor which satisfies
\begin{equation}
R(\theta)=R(-\theta)^{-1}=S(2\theta)R(i\pi-\theta)
\end{equation}
In the Lee-Yang theory there are two types of boundary conditions: 
\begin{itemize}
\item The identity boundary condition, denoted by $\mathbb{I}$, does not
have a parameter and its reflection factor is 
\begin{equation}
R(\theta)_{\mathbb{I}}=\left(1\right)_{\theta}\left(3\right)_{\theta}\left(-4\right)_{\theta}
\end{equation}
The pole at $i\frac{\pi}{2}$ 
\begin{equation}
R(\theta)_{\mathbb{I}}=\frac{g_{\mathbb{I}}^{2}}{2\theta-i\pi}+\mbox{reg.}\quad;\qquad g_{\mathbb{I}}=-2i\sqrt{(2\sqrt{3}-3)}\label{eq:Defg}
\end{equation}
shows that it can emit a virtual particle with zero energy but there
are no bound-states on this boundary.
\item The other type of boundary condition can accommodate a parameter,
$b$, as
\begin{equation}
R(\theta)_{\Phi}=R(\theta)_{\mathbb{I}}\left(b+1\right)_{\theta}\left(b-1\right)_{\theta}\left(5-b\right)_{\theta}\left(7-b\right)_{\theta}
\end{equation}
which can also emit a virtual zero energy particle with rate
\begin{equation}
g_{\Phi}\left(b\right)=\frac{\tan\left(\left(b+2\right)\frac{\pi}{12}\right)}{\tan\left(\left(b-2\right)\frac{\pi}{12}\right)}g_{\mathbb{I}}
\end{equation}
This boundary has bound states for $b>-1$, see \cite{Dorey:1998kt}
for the details. 
\end{itemize}
The scaling Lee-Yang theory is the only relevant perturbation of the
conformally invariant Lee-Yang model. This model has two conformal
invariant boundary conditions and the reflection factors correspond
to their perturbations. The identity conformal boundary has no relevant
operator as only the conformal descendents of the identity operator
can live on it. In contrast, the $\Phi$ conformal boundary additionally
contains the descendents of the boundary operator $\phi$ with weight
$h=-\frac{1}{5}$. In the perturbed theory the integrable boundary
perturbation introduces the parameter $b$, see \cite{Dorey:1998kt}
for the relation between the perturbed CFT and the scattering theory.

\subsection{Axioms}

Elementary boundary form factors are the matrix elements of local
boundary operators between the vacuum and asymptotic states:
\begin{equation}
F_{n}^{\mathcal{O}}(\theta_{1},\theta_{2},\dots,\theta_{n})=\langle0\vert\mathcal{O}(0)\vert\theta_{1},\theta_{2},\dots,\theta_{n}\rangle
\end{equation}
These form factors satisfy the following functional relations \cite{Bajnok:2006ze},
which are postulated as axioms%
\footnote{To streamline the notation we suppress the operator in the form factor,
if it does not lead to any confusion.%
}: 
\begin{equation}
F_{n}(\theta_{1},\dots,\theta_{i},\theta_{i+1},\dots,\theta_{n})=S(\theta_{i}-\theta_{i+1})F_{n}(\theta_{1},\dots,\theta_{i+1},\theta_{i},\dots,\theta_{n})\label{eq:axperm}
\end{equation}
\begin{equation}
F_{n}(\theta_{1},\dots,\theta_{n-1},\theta_{n})=R(\theta_{n})F_{n}(\theta_{1},\dots,\theta_{n-1},-\theta_{n})\label{eq:axrefl}
\end{equation}
\begin{equation}
F_{n}(\theta_{1},\theta_{2},\dots,\theta_{n})=R(i\pi-\theta_{1})F_{n}(2i\pi-\theta_{1},\theta_{2},\dots,\theta_{n})\label{eq:axcrossref}
\end{equation}
Additionally, they have singularities with residues, which are related
to form factors with less particles:
\begin{equation}
-i\mathop{\textrm{Res}}_{\theta=\theta'}F_{n+2}\left(\theta+i\pi,\theta',\theta_{1},\dots,\theta_{n}\right)=\biggr(1-\prod_{a=1}^{n}S\left(\theta-\theta_{a}\right)S\left(\theta+\theta_{a}\right)\biggr)F_{n}\left(\theta_{1},\dots,\theta_{n}\right)\label{eq:KinResAxiom}
\end{equation}
\begin{equation}
-i\mathop{\textrm{Res}}_{\theta=0}F_{n+1}\left(\theta+i\frac{\pi}{2},\theta_{1},\dots,\theta_{n}\right)=\frac{g}{2}\biggl(1-\prod_{a=1}^{n}S\left(i\frac{\pi}{2}-\theta_{a}\right)\biggr)F_{n}\left(\theta_{1},\dots,\theta_{n}\right)\label{eq:DynResAxiom}
\end{equation}

\begin{equation}
-i\mathop{\textrm{Res}}_{\theta=\theta'}F_{n+2}\left(\theta+i\frac{\pi}{3},\theta'-i\frac{\pi}{3},\theta_{1},\dots,\theta_{n}\right)=\Gamma F_{n+1}\left(\theta,\theta_{1},\dots,\theta_{n}\right)\label{eq:BdryResAxiom}
\end{equation}
Each solution of these functional relations with the appropriate asymptotic
properties corresponds to a local boundary operator of the theory
\cite{Szots:2007jq}. In the following we would like to present the
simplest explicit solutions. In doing so we introduce first a useful
parametrization.

\subsection{Parametrization of the form factors}

A parametrization, which fulfills the boundary form factor axioms
is given by \cite{Bajnok:2006ze}: 
\begin{equation}
F_{n}\left(\theta_{1},\theta_{2},\dots,\theta_{n}\right)=H_{n}Q_{n}\left(y_{1},y_{2},\dots,y_{n}\right)\prod_{i=1}^{n}\frac{r\left(\theta_{i}\right)}{y_{i}}\prod_{i<j}\frac{f\left(\theta_{i}-\theta_{j}\right)f\left(\theta_{i}+\theta_{j}\right)}{y_{i}+y_{j}},\label{eq:GenAnsatz}
\end{equation}
where $f(\theta)$ is the minimal solution of the bulk two particle
form factor equations
\begin{equation}
f(\theta)=S(\theta)f(-\theta)\qquad;\qquad f(i\pi-\theta)=f(i\pi+\theta)
\end{equation}
The function $r(\theta)$ is the minimal one particle boundary form
factor which satisfies 
\begin{equation}
r(\theta)=R(\theta)r(-\theta)\qquad;\qquad r(i\pi-\theta)=R(\theta)r(i\pi+\theta)
\end{equation}
and $Q_{n}$ is a symmetric polynomial of its arguments $y_{i}=e^{\theta_{i}}+e^{-\theta_{i}}$.
The polynomiality of $Q_{n}$ ensures for the correlation functions
in the conformal (short distance) limit to exhibit a polynomial separation
dependence. Finally, $H_{n}$ is some appropriately chosen normalization
constant proportional to the vacuum expectation value of the operator. 

In particular, for the scaling Lee-Yang model the minimal solution
of the bulk two particle form factor equation is
\begin{equation}
f\left(\theta\right)=\frac{y-2}{y+1}v\left(i\pi-\theta\right)v\left(-i\pi+\theta\right)\quad;\qquad y=e^{\theta}+e^{-\theta}
\end{equation}
where
\begin{equation}
\log v\left(\theta\right)=2\int_{0}^{\infty}\frac{dt}{t}\frac{\sinh\frac{t}{2}\sinh\frac{t}{3}\sinh\frac{t}{6}}{\sinh^{2}t}e^{i\frac{\theta t}{\pi}}
\end{equation}
The minimal one particle form factor depends on the boundary condition.
For the identity boundary
\begin{equation}
r_{\mathbb{I}}\left(\theta\right)=4iu(\theta)\sinh\theta
\end{equation}
with 
\begin{equation}
\log u(\theta)=\int_{0}^{\infty}\frac{dt}{t}\frac{\sinh t-\cos\left(\left(\frac{i}{2}-\frac{\theta}{\pi}\right)t\right)\left(\sinh\frac{5t}{6}+\sinh\frac{t}{2}-\sinh\frac{t}{3}\right)}{\sinh\frac{t}{2}\sinh t}
\end{equation}
and a useful normalization is $H_{n}=\langle\mathcal{O}\rangle\left(\frac{i\sqrt[4]{3}}{v\left(0\right)\sqrt{2}}\right)^{n}$. 

With these choices the form factor axioms Eqs. (\ref{eq:axperm}-\ref{eq:axcrossref})
are automatically satisfied, while the singularity axioms provide
recursion relations for the polynomials $Q_{n}$:
\begin{eqnarray}
Q_{n+2}\left(y_{+},y_{-},y_{1},\dots,y_{n}\right) & = & D_{n}\left(y\vert y_{1},\dots,y_{n}\right)Q_{n+1}\left(y,y_{1},\dots,y_{n}\right)\label{eq:DynRec}\\
Q_{n+2}\left(y,-y,y_{1},\dots,y_{n}\right) & = & P_{n}\left(y\vert y_{1},\dots,y_{n}\right)Q_{n}\left(y_{1},\dots,y_{n}\right)\label{eq:KinRec}\\
Q_{n+1}\left(0,y_{1},\dots,y_{n}\right) & = & B_{n}\left(y_{1},\dots,y_{n}\right)Q_{n}\left(y_{1},\dots,y_{n}\right)\label{eq:BdryRec}
\end{eqnarray}
with
\begin{eqnarray}
D_{n}\left(y\vert y_{1},\dots,y_{n}\right) & = & \prod_{i=1}^{n}\left(y+y_{i}\right)\label{eq:DynPoly}\\
P_{n}\left(y\vert y_{1},\dots,y_{n}\right) & = & \frac{\prod_{i=1}^{n}\left(y_{i}-y_{-}\right)\left(y_{i}+y_{+}\right)-\prod_{i=1}^{n}\left(y_{i}+y_{-}\right)\left(y_{i}-y_{+}\right)}{2\left(y_{+}-y_{-}\right)}\label{eq:KinPoly}\\
B_{n}\left(y_{1},\dots,y_{n}\right) & = & \frac{\prod_{i=1}^{n}\left(y_{i}+\sqrt{3}\right)-\prod_{i=1}^{n}\left(y_{i}-\sqrt{3}\right)}{2\sqrt{3}}\label{eq:BdryPoly}
\end{eqnarray}
Here we introduced $y_{\pm}=2\cosh(\theta\pm i\frac{\pi}{3})$. In
the next section we explicitly construct a polynomial solution of
these recursion relations.  

Let us summarize the similar parametrization for the $\Phi$ boundary.
We distinguish the form factors and polynomials from those of the
identity boundary by a tilde: 
\begin{equation}
\tilde{F}_{n}\left(\theta_{1},\dots,\theta_{n}\right)=H_{n}\prod_{i=1}^{n}\frac{r_{\Phi}\left(\theta_{i}\right)}{y_{i}}\prod_{i<j}\frac{f\left(\theta_{1}-\theta_{j}\right)f\left(\theta_{i}+\theta_{j}\right)}{y_{i}+y_{j}}\tilde{Q}_{n}\left(y_{1},\dots,y_{n}\right)\label{eq:PhiAnsatz}
\end{equation}
Here the one particle form factor has poles corresponding to the possible
bound-states and takes the form
\begin{equation}
r_{\Phi}\left(\theta\right)=\frac{i\sinh\theta}{\left(\sinh\theta-i\sin\left(\left(b-1\right)\frac{\pi}{6}\right)\right)\left(\sinh\theta-i\sin\left(\left(b+1\right)\frac{\pi}{6}\right)\right)}u\left(\theta\right)
\end{equation}
Due to this factor the recursion relations are slightly modified:
\begin{eqnarray}
\tilde{Q}_{n+2}\left(y_{+},y_{-},y_{1},\dots,y_{n}\right) & = & \left(y^{2}-3+\alpha\right)D_{n}\left(y\vert y_{1},\dots,y_{n}\right)\tilde{Q}_{n+1}\left(y,y_{1},\dots,y_{n}\right)\label{eq:DynRecPhi}\\
\tilde{Q}_{n+2}\left(y,-y,y_{1},\dots,y_{n}\right) & = & \left(y^{4}-\left(3+\alpha\right)y^{2}+\alpha^{2}\right)P_{n}\left(y\vert y_{1},\dots,y_{n}\right)\tilde{Q}_{n}\left(y_{1},\dots,y_{n}\right)\qquad\label{eq:KinRecPhi}\\
\tilde{Q}_{n+1}\left(0,y_{1},\dots,y_{n}\right) & = & \alpha B_{n}\left(y_{1},\dots,y_{n}\right)\tilde{Q}_{n}\left(y_{1},\dots,y_{n}\right)\label{eq:BdryRecPhi}
\end{eqnarray}
where $\alpha=1+2\cos\frac{b\pi}{3}$. 

Finally we note that the solutions of the form factor axioms and the
space of operators are related. We call the tower of form factor solutions\emph{
}\textit{\emph{\cite{Koubek:1993ke,Szots:2007jq}}} the set $\mathcal{F}_{\mathcal{O}}=\left\{ F_{n}^{\mathcal{O}}\left(\theta_{1},\dots,\theta_{n}\right)\right\} _{n\in\mathbb{N}}$,
which satisfies the form factor axioms (\ref{eq:axperm}-\ref{eq:BdryResAxiom}).
It was pointed out for the bulk Lee-Yang model \cite{Koubek:1994zp,Koubek:1994gk}
and also for the boundary case \cite{Szots:2007jq} that there is
a one-to-one correspondence between these towers and the boundary
operator content of the ultraviolet conformal field theory.

Every such tower starts with a so-called kernel solution. An $n^{\mathrm{th}}$
level kernel solution is defined as a polynomial of $n$ variables\emph{
}whose value is zero at each pole\emph{ }and in case of the boundary
Lee-Yang model is given as \cite{Szots:2007jq}
\begin{equation}
\sigma_{k_{1}}^{\left(n\right)}\dots\sigma_{k_{l}}^{\left(n\right)}\prod_{1\leq i<j\leq n}\left(y_{i}+y_{j}\right)\prod_{1\leq i<j\leq n}\left(y_{i}^{2}+y_{i}y_{j}+y_{j}^{2}-3\right)\prod_{i=1}^{n}y_{i}
\end{equation}
where $0<k_{1}\leq k_{2}\leq\dots\leq k_{l}\leq n$.

\section{Explicit Boundary Form Factor Solutions\label{sec:Explicit-Boundary-Form}}

In this section we explicitly solve the recurrence relations for operators
with the mildest ultraviolet behavior. As $Q_{n}$ and $\tilde{Q}_{n}$
are symmetric polynomials we introduce the following basis of homogeneous
symmetric polynomials: 
\begin{equation}
\prod_{i=1}^{n}(y+y_{i})=\sum_{k}y^{n-k}\sigma_{k}^{(n)}(y_{1},\dots,y_{n})
\end{equation}
With this definition $\sigma_{k}^{(n)}=0$ if $k<0$ or $k>n$.

\subsection{The Identity Boundary}

In this subsection we solve explicitly the recurrence relations (\ref{eq:DynRec}-\ref{eq:BdryRec})
for the operator which has the mildest ultraviolet behavior, i.~e.
the off-critical version of the boundary stress-tensor. This is the
tower of form factors built over the first level kernel solution $\sigma_{1}$\emph{
}\cite{Bajnok:2006ze}\emph{.}

\subsubsection{Formulating the conjecture\label{sub:Formulating-the-conjecture}}

The lowest lying solutions are given as \cite{Szots:2007jq} 
\begin{equation}
Q_{1}^{T}=\sigma_{1}\quad,\qquad Q_{2}^{T}=\sigma_{1}\quad,\qquad Q_{3}^{T}=\sigma_{1}^{2}\quad,\qquad Q_{4}^{T}=\sigma_{1}^{2}\left(\sigma_{2}+3\right).
\end{equation}
As the $\sigma_{1}$ symmetric polynomial has the properties
\begin{eqnarray}
\sigma_{1}^{\left(n+2\right)}\left(y,-y,y_{1},\dots,y_{n}\right) & = & \sigma_{1}^{\left(n\right)}\left(y_{1},\dots,y_{n}\right)\nonumber \\
\sigma_{1}^{\left(n+2\right)}\left(y_{+},y_{-},y_{1},\dots,y_{n}\right) & = & \sigma_{1}^{\left(n+1\right)}\left(y,y_{1},\dots,y_{n}\right)\nonumber \\
\sigma_{1}^{\left(n+1\right)}\left(0,y_{1},\dots,y_{n}\right) & = & \sigma_{1}^{\left(n\right)}\left(y_{1},\dots,y_{n}\right)
\end{eqnarray}
 the tower $Q_{n}^{T}\sigma_{1}^{k}$ for $k>0$ will also satisfy
the recursion. This tower is claimed to correspond to the operator
$\partial^{k}T$ \cite{Szots:2007jq}.

In what follows we give explicit formulae for these $Q_{n}^{T}$ polynomials.
Even if our method is special for the Lee-Yang model and demand some
guesswork one can expect that similar method adapted to some other
integrable models could also work.

We start with the leading order analysis. Let us take all the rapidities
and shift them uniformly $\theta_{i}\rightarrow\theta_{i}+\Lambda$.
The $Q_{n}$ polynomials are functions of the variables $y_{i}=x_{i}+x_{i}^{-1}$
with $x_{i}=e^{\theta_{i}},$ therefore they are large if $\Lambda\rightarrow\pm\infty$.
The recursion polynomials can then be expanded in powers of $\lambda=e^{\Lambda}$
and one finds
\begin{eqnarray}
D_{n}\left(y\vert y_{1},\dots,y_{n}\right) & \sim & \lambda^{\pm n}V_{n}\left(x^{\pm1}\vert x_{1}^{\pm1},\dots,x_{n}^{\pm1}\right)\nonumber \\
P_{n}\left(y\vert y_{1},\dots,y_{n}\right) & \sim & \lambda^{\pm\left(2n-1\right)}\left(-1\right)^{n+1}U_{n}\left(x^{\pm1}\vert x_{1}^{\pm1},\dots,x_{n}^{\pm1}\right)\quad;\qquad\Lambda\rightarrow\pm\infty\nonumber \\
B_{n}\left(y_{1},\dots,y_{n}\right) & \sim & \lambda^{\pm n}0
\end{eqnarray}
where
\begin{eqnarray}
U_{n}\left(x\vert x_{1},\dots,x_{n}\right) & = & \frac{\prod_{i=1}^{n}\left(x+\omega x_{i}\right)\left(x-\omega^{-1}x_{i}\right)-\prod_{i=1}^{n}\left(x-\omega x_{i}\right)\left(x+\omega^{-1}x_{i}\right)}{2x\left(\omega-\omega^{-1}\right)}\nonumber \\
V_{n}\left(x\vert x_{1},\dots,x_{n}\right) & = & \prod_{i=1}^{n}\left(x+x_{i}\right)
\end{eqnarray}
are the polynomials appearing in the bulk recursion \cite{Zamolodchikov:1990bk}.
Here we introduced $\omega=e^{\frac{i\pi}{3}}.$ Its solution is given
in a determinant form 
\begin{eqnarray}
\det\Sigma\left(x,-x,x_{1},\dots,x_{n}\right) & = & \left(-1\right)^{n+1}U_{n}\left(x\vert x_{1},\dots,x_{n}\right)\det\Sigma\left(x_{1},\dots,x_{n}\right)\nonumber \\
\det\Sigma\left(\omega x,\omega^{-1}x,x_{1},\dots x_{n}\right) & = & V_{n}\left(x\vert,x_{1},\dots,x_{n}\right)\det\Sigma\left(x,x_{1},\dots,x_{n}\right)
\end{eqnarray}
for $n\geq4$ where the $\Sigma$ matrix is given as $\Sigma_{i,j}=\sigma_{3j-2i+1}$.
Here we need to comment on the notations. The $\Sigma$ matrix is
expressed in terms of the $\sigma_{k}$ symmetric polynomials and
its form is universal for all $n$. In what follows we  use the following
abbreviation: if the arguments are not written explicitly we  denote
by $\Sigma\left(n\right)$ the matrix with $n$ general entries. The
$n$ dependence appears through the size of the matrix and also comes
from the range of the elementary symmetric polynomials as $\sigma_{k}^{\left(n\right)}=0$
if $k>n$, thus we have
\begin{equation}
\Sigma\left(n\right)_{i,j}=\sigma_{3j-2i+1}^{\left(n\right)}\quad,\qquad1\leq i,\, j\leq n-3
\end{equation}

In the bulk case the determinant form is motivated by the clustering
property of the form factors \cite{Koubek:1993ke}. For the boundary
problem the simplest idea is to try to find some lower order corrections
to the $\Sigma$ matrix and interestingly, even though the clustering
argument fails we managed to obtain the matrix explicitly.

One can determine the $Q_{n}^{T}$ polynomials for small $n$ and
then organize them into a determinant
\begin{equation}
Q_{n}^{T}\left(y_{1},\dots,y_{n}\right)=\sigma_{1}^{2}\left(y_{1},\dots,y_{n}\right)\det\Xi\left(y_{1},\dots,y_{n}\right)\quad,\qquad n\geq4\label{eq:IdFFform}
\end{equation}
where the first few are
\[
\Xi\left(4\right)=\left(\sigma_{2}^{\left(4\right)}+3\right)\qquad;\qquad\Xi\left(5\right)=\left(\begin{array}{cc}
\sigma_{2}^{\left(5\right)}+3 & \sigma_{5}^{\left(5\right)}\\
1 & 3\sigma_{1}^{\left(5\right)}+\sigma_{3}^{\left(5\right)}
\end{array}\right)
\]
\[
\Xi\left(6\right)=\left(\begin{array}{ccc}
\sigma_{2}^{\left(6\right)}+3 & \sigma_{5}^{\left(6\right)} & -3\sigma_{6}^{\left(6\right)}\\
1 & 3\sigma_{1}^{\left(6\right)}+\sigma_{3}^{\left(6\right)} & \sigma_{6}^{\left(6\right)}\\
0 & \sigma_{1}^{\left(6\right)} & 3\sigma_{2}^{\left(6\right)}+\sigma_{4}^{\left(6\right)}+9
\end{array}\right)
\]
\[
\Xi\left(7\right)=\left(\begin{array}{cccc}
\sigma_{2}^{\left(7\right)}+3 & \sigma_{5}^{\left(7\right)} & -3\sigma_{6}^{\left(7\right)} & 9\sigma_{7}^{\left(7\right)}\\
1 & 3\sigma_{1}^{\left(7\right)}+\sigma_{3}^{\left(7\right)} & \sigma_{6}^{\left(7\right)} & -3\sigma_{7}^{\left(7\right)}\\
0 & \sigma_{1}^{\left(7\right)} & 3\sigma_{2}^{\left(7\right)}+\sigma_{4}^{\left(7\right)}+9 & \sigma_{7}^{\left(7\right)}\\
0 & 0 & \sigma_{2}^{\left(7\right)}+6 & 9\sigma_{1}^{\left(7\right)}+3\sigma_{3}^{\left(7\right)}+\sigma_{5}^{\left(7\right)}
\end{array}\right)
\]
and we adapted the same notations as we explained for $\Sigma$. We
determined the form factors and packed them into similar matrix forms
up to $n=16$. 

These matrices already suggest some properties for the $\Xi$ matrix.
We can see that at least for small $n$ every term in the last column
of $\Xi\left(n\right)$ are proportional to $\sigma_{n}^{\left(n\right)}$
except the right bottom corner, which does not contain $\sigma_{n}^{\left(n\right)}$
at all. Suppose that this is the case and let us analyze the boundary
kinematical relation (\ref{eq:BdryRec}). As the symmetric polynomials
have the properties $\sigma_{k}^{\left(n+1\right)}\left(0,y_{1},\dots,y_{n}\right)=\sigma_{k}^{\left(n\right)}\left(y_{1},\dots,y_{n}\right)$
if $k<n+1$ and $\sigma_{n+1}^{\left(n+1\right)}\left(0,y_{1},\dots,y_{n}\right)=0$
we have
\begin{equation}
\Xi\left(0,y_{1},\dots,y_{n}\right)=\left(\begin{array}{ccc|c}
 &  &  & 0\\*
 & \Xi\left(y_{1},\dots,y_{n}\right) &  & \vdots\\
 &  &  & 0\\
\hline * & \dots & * & B_{n}
\end{array}\right)
\end{equation}
Expanding the determinant along the last column we can conclude that
the element on the right bottom corner is the boundary polynomial
$B_{n}$ thus these are the diagonal elements of $\Xi$.

A uniform shift $\theta_{k}\rightarrow\theta_{k}+i\pi$ leads to the
transformation $y_{k}\rightarrow-y_{k}$. Apparently as in the bulk
case the $Q_{n}^{T}$ polynomials have a definite parity under this
transformation which is consistent with the fact that the recursion
polynomials do also have a definite and compatible parity. This can
be reached if in $\Xi$ every matrix element has a definite parity
thus it contains only odd or only even $\sigma$ polynomials.

Finally, we can observe that if the correction to a specific matrix
element has the form of 
\begin{equation}
\sigma_{k}+a_{1}\sigma_{k-2}+a_{2}\sigma_{k-4}\dots=\sigma_{k}+\sum_{j=1}a_{j}\sigma_{k-2j}
\end{equation}
where $\sigma_{k}$ is the leading order term, then the coefficient
can be written as $a_{j}=3^{j}b_{j}$ with some integer $b_{j}$.
Now if we arrange the $b_{j}$ numbers to a table (up to $n=16$)
one can recognize the elements of Pascal's triangle. Then if we define
\begin{equation}
\binom{m}{k}=\begin{cases}
\frac{m\left(m-1\right)\left(m-2\right)\dots\left(m-k+1\right)}{k!}\qquad & \mathrm{if}\ k\in\mathbb{N}\\
0 & \mathrm{otherwise}
\end{cases}
\end{equation}
we can formulate our conjecture, namely
\begin{equation}
\Xi\left(n\right)_{i,j}=\sum_{k\in\mathbb{Z}}3^{k}\tbinom{i-j+k}{k}\sigma_{3j-2i+1-2k}^{\left(n\right)}\qquad1\leq i,j\leq n-3\label{eq:XiConj}
\end{equation}
We prove in the Appendix, that (\ref{eq:IdFFform}) indeed satisfies
all the recurrence relations (\ref{eq:DynRec}- \ref{eq:BdryRec}).

\subsection{The $\Phi$ Boundary}

In this subsection we use the fusion method to extend the previous
results for the $\Phi$-boundary. We then determine the form factors
of the boundary primary operator $\phi$, which lives only on this
boundary.

\subsubsection{Fusion method}

From the perturbed CFT point of view it follows that the space of
operators living on the identity boundary is the Verma module built
over the conformal vacuum, while the space of operators living on
the $\Phi$ boundary is the direct sum of the Verma module of the
conformal vacuum and the Verma module built over the only other highest
weight conformal vector $\phi$ of weight $-\frac{1}{5}$. 

In \cite{Bajnok:2007jg} it was proven that the $\Phi$-boundary can
be thought of if the only non-trivial defect (the $\Phi$-defect)
were fused to the identity boundary. In the language of the form factor
solutions \cite{Bajnok:2009hp} this means that the form factors of
the operators which are present on both boundaries (i.~e. operators
form the vacuum module) are related as%
\footnote{In the CFT limit the defect is not seen by the energy momentum tensor,
i.e. it is continuous there.%
}
\begin{equation}
\tilde{F}_{n}\left(\theta_{1},\dots,\theta_{n}\right)=\prod_{i=1}^{n}T_{-}\left(\theta_{i}\right)F_{n}\left(\theta_{1},\dots,\theta_{n}\right)\label{eq:fusion_FF}
\end{equation}
where $T_{-}$ is the defect transmission factor which is 
\begin{equation}
T_{-}\left(\theta\right)=S\left(\theta-i(3-b)\frac{\pi}{6}\right)=-\frac{\sinh\left(\frac{\theta}{2}+\left(b+1\right)\frac{i\pi}{12}\right)}{\sinh\left(\frac{\theta}{2}+\left(b-5\right)\frac{i\pi}{12}\right)}\frac{\sinh\left(\frac{\theta}{2}+\left(b-1\right)\frac{i\pi}{12}\right)}{\sinh\left(\frac{\theta}{2}+\left(b-7\right)\frac{i\pi}{12}\right)}.
\end{equation}
Here $b$ is the defect parameter which, after fusion, becomes the
parameter of the $\Phi$-boundary. Plugging back the Ansatz (\ref{eq:GenAnsatz})
and (\ref{eq:PhiAnsatz}) to (\ref{eq:fusion_FF}) we get relations
between the polynomials $Q$ and $\tilde{Q}$, namely
\begin{eqnarray}
\tilde{Q}_{n}\left(y_{1},\dots,y_{n}\right) & = & \left(\prod_{i=1}^{n}T_{-}\left(\theta_{i}\right)\frac{r_{\mathbb{I}}\left(\theta_{i}\right)}{r_{\Phi}\left(\theta_{i}\right)}\right)Q_{n}\left(y_{1},\dots,y_{n}\right)\nonumber \\
 & = & \prod_{i=1}^{n}\left(\alpha-\sqrt{3}\sqrt{\alpha+1}y_{i}+y_{i}^{2}\right)Q_{n}\left(y_{1},\dots,y_{n}\right)
\end{eqnarray}
It is straightforward to check that if $Q_{n}$ satisfies the equations
(\ref{eq:DynRec}-\ref{eq:BdryRec}) then $\tilde{Q}_{n}$ indeed
satisfies (\ref{eq:DynRecPhi}-\ref{eq:BdryRecPhi}).

\subsubsection{Form Factors of the boundary primary field}

At the conformal point the $\Phi$-boundary contains also operators
from the module of the conformal primary field $\phi$. In \cite{Szots:2007jq}
it was argued by a counting argument that there is a one-to-one correspondence
between the conformal boundary fields and the solutions of the form
factor equations. It was also argued that the form factor solution
with the mildest ultraviolet behavior corresponds to the primary field
$\phi$. The first few solutions of the $\Phi$-boundary recurrence
relations for $\phi$ are
\begin{eqnarray}
 & \tilde{Q}_{1}^{\phi}=\sigma_{1}\quad,\qquad\tilde{Q}_{2}^{\phi}=\sigma_{1}\left(\sigma_{2}+\alpha\right)\quad,\qquad\tilde{Q}_{3}^{\phi}=\sigma_{1}\left(\alpha\sigma_{1}\left(\alpha+\sigma_{2}\right)+\left(\sigma_{2}+3\right)\sigma_{3}\right)\nonumber \\
 & \tilde{Q}_{4}^{\phi}=\alpha\sigma_{1}\left(\sigma_{2}+3\right)\left(\sigma_{1}\left(\alpha^{2}+\alpha\sigma_{2}\right)+\left(\sigma_{2}+3\right)\sigma_{3}\right)+\sigma_{1}\left(\sigma_{2}+3\right)\left(3\sigma_{1}+\sigma_{3}\right)\sigma_{4}\label{eq:FirstSolutionsPhi}
\end{eqnarray}
We attempt to determine the whole tower of solutions based on these
first members (\ref{eq:FirstSolutionsPhi}). To this end we take a
similar Ansatz as was proposed for the defect case \cite{Bajnok:2013eaa},
namely 
\begin{equation}
\tilde{Q}_{n}^{\phi}=\sigma_{1}^{\left(n\right)}S_{n}\det\Xi\left(n\right)\qquad n\geq4
\end{equation}
The $S_{n}$ polynomials are defined only for $n\geq4$ and have to
satisfy the recurrence relations
\begin{eqnarray}
S_{n+2}\left(y_{+},y_{-},y_{1},\dots,y_{n}\right) & = & \left(\alpha-3+y^{2}\right)S_{n+1}\left(y,y_{1},\dots,y_{n}\right)\nonumber \\
S_{n+2}\left(y,-y,y_{1},\dots,y_{n}\right) & = & \left(y^{4}-\left(3+\alpha\right)y^{2}+\alpha^{2}\right)S_{n}\left(y_{1},\dots,y_{n}\right)\nonumber \\
S_{n+1}\left(0,y_{1},\dots,y_{n}\right) & = & \alpha S_{n}\left(y_{1},\dots,y_{n}\right)\label{eq:S_rec}
\end{eqnarray}
When it does not lead to any confusion we do not write out explicitly
the arguments of $S_{n}$ keeping in mind that $S_{n}$ always has
$n$ arguments. We can compute explicitly the first few solutions which
can be cast to the form
\begin{equation}
S_{n}=\sum_{k\in\mathbb{Z}}p_{k}\left(\alpha\right)\kappa_{n-2k-1}^{(n)}\label{eq:SAnsatz}
\end{equation}
where we introduced the 
\begin{equation}
\kappa_{k}^{\left(n\right)}=\sum_{l=0}^{k}\alpha^{l}\sigma_{n-l}^{\left(n\right)}\sigma_{k-l}^{\left(n\right)}=\sum_{l\in\mathbb{Z}}\alpha^{l}\sigma_{n-l}^{\left(n\right)}\sigma_{k-l}^{\left(n\right)}\label{eq:KappaDef}
\end{equation}
polynomials. Here in the last equation we used the fact that the symmetric
polynomials are defined to be zero whenever their index is negative
or have less arguments then their index. These polynomials have the
properties
\begin{eqnarray}
\kappa_{k}^{(n+2)}\left(y_{+},y_{-},y_{1},...,y_{n}\right) & = & \alpha^{2}\kappa_{k-2}^{(n)}+\alpha^{2}y\kappa_{k-3}^{(n)}+\alpha^{2}(y^{2}-3)\kappa_{k-4}^{(n)}+\alpha y\kappa_{k-1}^{(n)}+\alpha y^{2}\kappa_{k-2}^{(n)}+\nonumber \\
 &  & \alpha y(y^{2}-3)\kappa_{k-3}^{(n)}+(y^{2}-3)\kappa_{k}^{(n)}+y(y^{2}-3)\kappa_{k-1}^{(n)}+(y^{2}-3)^{2}\kappa_{k-2}^{(n)}\nonumber \\
\kappa_{k}^{(n+1)}\left(y,y_{1},...,y_{n}\right) & = & \alpha\kappa_{k-1}^{(n)}+\alpha y\kappa_{k-2}^{(n)}+y\kappa_{k}^{(n)}+y^{2}\kappa_{k-1}^{(n)}\nonumber \\
\kappa_{k}^{(n+2)}\left(y,-y,y_{1},...,y_{n}\right) & = & \alpha^{2}\kappa_{k-2}^{(n)}-y^{2}\alpha^{2}\kappa_{k-4}^{(n)}-y^{2}\kappa_{k}^{(n)}+y^{4}\kappa_{k-2}^{(n)}\nonumber \\
\kappa_{k}^{(n+1)}\left(0,y_{1},...,y_{n}\right) & = & \alpha\kappa_{k-1}^{(n)}\label{eq:KappaPropty}
\end{eqnarray}
where on the right hand side of (\ref{eq:KappaPropty}) the arguments
of every $\kappa^{\left(n\right)}$ are taken to be $\left(y_{1},\dots,y_{n}\right)$.
The functions $p_{k}\left(\alpha\right)$ appearing on the right hand
side of eq. (\ref{eq:SAnsatz}) are polynomials of $\alpha$ and the
first few are
\begin{eqnarray}
p_{0}\left(\alpha\right)=1 & ; & p_{1}\left(\alpha\right)=3\\
p_{2}\left(\alpha\right)=9+3\alpha-\alpha^{2} & \quad;\quad & p_{3}\left(\alpha\right)=27+18\alpha-3\alpha^{2}-\alpha^{3}\nonumber 
\end{eqnarray}
For negative indices they are defined to be zero, $p_{k}=0$ if $k<0$,
which means that the sum on the right hand side of (\ref{eq:SAnsatz})
is finite.

Plugging back the Ansatz (\ref{eq:SAnsatz}) to the recursion (\ref{eq:S_rec})
and taking advantage of the identities (\ref{eq:KappaPropty}) one
can easily derive that the recursion (\ref{eq:S_rec}) is satisfied
provided
\begin{equation}
p_{k+1}\left(\alpha\right)-\left(\alpha+3\right)p_{k}\left(\alpha\right)+\alpha^{2}p_{k-1}\left(\alpha\right)=0\qquad k\geq1.\label{eq:pRec}
\end{equation}
This is an ordinary second order recursion which can be solved by
usual techniques. 

The final solution for the $p_{k}$ polynomials has a simpler form
in terms of the boundary parameter $b$ :
\[
p_{k}\left(b\right)=C\left(b\right)r\left(b\right)^{k}+D\left(b\right)q\left(b\right)^{k}
\]
with
\begin{eqnarray}
r\left(b\right)=4\cos^{2}\left((b+1)\frac{\pi}{6}\right) & \quad;\qquad & q\left(b\right)=4\cos^{2}\left((b-1)\frac{\pi}{6}\right)\nonumber \\
C\left(b\right)=\frac{1}{\sqrt{3}}\frac{\cos\frac{(b+1)\pi}{6}}{\cos\frac{b\pi}{6}} & \quad;\qquad & D\left(b\right)=\frac{1}{\sqrt{3}}\frac{\cos\frac{(b-1)\pi}{6}}{\cos\frac{b\pi}{6}}
\end{eqnarray}
Note that if $b\in6\mathbb{Z}-3$ the polynomial $p_{k}$ is simply
the limit of the above formulae which is $p_{k}=1+2k$.

\section{Conclusion\label{sec:Conclusion}}

In this paper we gave explicit closed formulae for the form factors
of the boundary fields with the lowest scaling dimensions in the scaling
Lee-Yang model for all integrable boundary conditions. We first determined
the generic $n$-particle form factor of the energy-momentum tensor
for the identity boundary in a determinant form. We then applied the
fusion idea to derive the corresponding $n$-particle form factors
in case of the $\Phi$-boundary and proved the consistency of the
solutions. Finally, based on our experience with defect form factors,
we presented the only remaining form factors for the boundary primary
field $\phi$. We emphasize these results are the first explicit \emph{boundary}
form factor solutions. 

It is very remarkable that the form factor solution we found takes
a determinant form. Actually in finding the solution we exploited
the fact that for large rapidities the boundary form factor reduces
to the bulk one, which took already a determinant form. We then systematically
determined the lower order entries of this determinant. We expect
that similar method can work for other models including the boundary
form factors of the exponential operators in the sinh-Gordon and sine-Gordon
models \cite{Takacs:2008je,Lencses:2011ab}. 

The explicit form of the form factors are very useful. They can be
used to calculate correlation functions in infinite and also in finite
volumes \cite{Kormos:2007qx} or describe the exact finite volume/temperature
boundary vacuum expectation values following \cite{Takacs:2008ec}.

\section*{Acknowledgments}

We thank Gábor Takacs and Máté Lencsés for the useful discussions.
We were supported by the Lendület Grant LP2012-18/2014 and by OTKA 81461. 

\appendix

\section{Proof of the conjecture\label{sec:Proof-of-the}}

Throughout the appendix we do not display twice the number of arguments
of any object. This means for an elementary symmetric polynomial that
we use the new notation $\sigma_{k}^{\left(n\right)}\equiv\sigma_{k}(y_{1},\dots,y_{n})$.
Let us recall our notations: The $\Xi$ matrix is considered to be
an infinite matrix and expressed in terms of the $\sigma_{k}$ symmetric
polynomials. Its form is universal for all $n$. If we consider the
matrix $\Xi$ with $n$ general arguments, then we cut the infinite
matrix into an $\left(n-3\right)\times\left(n-3\right)$ submatrix
and denote it by $\Xi\left(n\right)$. The $n$-dependence appears
through the size of the matrix and also comes from the range of the
elementary symmetric polynomials as $\sigma_{k}^{\left(n\right)}=0$
if $k>n$, see (\ref{eq:XiConj}). During the proof we  use the notation
\begin{equation}
\Xi_{i,j}^{\left(s\right)}=\sum_{k\in\mathbb{Z}}3^{k}\tbinom{i-j+k}{k}\sigma_{3j-2i+1-2k-s}\label{eq:XiShifted}
\end{equation}
and also $\Xi^{\left(s\right)}\left(n\right)$ with similar conventions.

In this appendix we also consider the recursion polynomials as functions
of the elementary symmetric polynomials:
\begin{eqnarray}
D_{n} & = & \sum_{i}y^{n-i}\sigma_{i}\label{eq:Dsigma}\\
P_{n} & = & \sum_{q<p}\mathcal{P}_{p,q}\sigma_{n-p}\sigma_{n-q}\label{eq:Psigma}\\
B_{n} & = & \sum_{k}3^{k}\sigma_{n-1-2k}\label{eq:Bsigma}
\end{eqnarray}
where
\[
\mathcal{P}_{p,q}=\frac{y_{+}^{p}y_{-}^{q}\left((-1)^{q}-(-1)^{p}\right)+y_{+}^{q}y_{-}^{p}\left((-1)^{p}-(-1)^{q}\right)}{2\left(y_{+}-y_{-}\right)}
\]
When these quantities are taken at specified arguments we  use similar
abbreviation as for $\Xi$. Note that these functions reduce to the
recursion polynomials (\ref{eq:DynPoly}-\ref{eq:BdryPoly}) only
if the appropriate number of arguments is chosen, for example $D_{n}\left(y\vert y_{1},\dots,y_{n}\right)=D_{n}\left(n\right)$.

\subsection{General ideas}

Since the solution of the recursion equations (\ref{eq:DynRec}, \ref{eq:KinRec},
\ref{eq:BdryRec}) is given in a determinant form (without the factor
$\sigma_{1}^{2}$) the proof of the conjecture should reflect this
property. Namely, we first expand the elements of the determinants
with special arguments on the left hand side of these equations by
exploiting the properties of elementary symmetrical polynomials:
\begin{eqnarray}
\sigma_{k}\left(y,-y,y_{1},\dots,y_{n}\right) & = & \sigma_{k}^{(n)}-y^{2}\sigma_{k-2}^{(n)}\label{eq:A1}\\
\sigma_{k}\left(y_{+},y_{-},y_{1},\dots,y_{n}\right) & = & \sigma_{k}^{(n)}+y\sigma_{k-1}^{(n)}+\left(y^{2}-3\right)\sigma_{k-2}^{(n)}\label{eq:A2}\\
\sigma_{k}\left(0,y_{1},\dots,y_{n}\right) & = & \sigma_{k}^{(n)}\label{eq:A3}
\end{eqnarray}
We then manipulate the rows and columns systematically, until they
get into a form in which the equations hold true explicitly. In the
following paragraphs we  present these desired forms and the manipulation
algorithms.

The exact proofs would get rather technical later on, therefore in
these cases only brief outlines are given. During the proof we use
the fact that the determinant of a matrix does not change if we add
to a row (column) any scalar time an other row (column) and by such
steps we  reduce the matrices to a special form.

\subsection{Boundary recursion}

By construction of the matrix (\ref{eq:XiConj}) the boundary recursion
equation holds trivially, see Subsection \ref{sub:Formulating-the-conjecture}.

\subsection{Dynamical recursion\label{sub:Dynamical-recursion}}

The desired reduced form of $\Xi\left(y_{+},y_{-},y_{1},\dots,y_{n}\right)$
after expanding its elementary symmetric polynomials using (\ref{eq:A2})
is 
\begin{equation}
\left(\begin{array}{ccc|c}
 &  &  & 0\\*
 & \Xi\left(y,y_{1},\dots,y_{n}\right) &  & \vdots\\
 &  &  & 0\\
\hline * & \dots & * & D_{n}
\end{array}\right)
\end{equation}
where the argument of $D_{n}$ is $D_{n}\left(y\vert y_{1},\dots,y_{n}\right)$.
This is so because after calculating the determinant by expanding
it along the last column only $D_{n}\left(y\vert y_{1},\dots,y_{n}\right)\det(\Xi\left(y,y_{1},\dots,y_{n}\right))$
remains, which is exactly what the dynamical recursion requires.

The algorithm consists of two steps, whose order is arbitrary. During
the column operation we add recursively from left to right each column
times $y(y^{2}-3)$ to the one to its right. The row operations consist
of adding from top to bottom to each row $\left(3-y^{2}\right)$ times
the one below it, (i.~e. not recursively). Note that while the column
operation results in a cumulative sum the row operation does not. 

The expansion of the elementary symmetric polynomials in $\Xi\left(y_{+},y_{-},y_{1},\dots,y_{n}\right)$
can be written as 
\begin{equation}
\Xi\left(y_{+},y_{-},y_{1},\dots,y_{n}\right)=\Xi^{(0)}(n)+y\Xi^{(1)}(n)+(y^{2}-3)\Xi^{(2)}(n)
\end{equation}
After the row operation the $(i,j)$ element of our matrix becomes
\begin{equation}
\Xi_{i,j}^{(0)}+y\Xi_{i,j}^{(1)}+(y^{2}-3)(\Xi_{i,j}^{(2)}-\Xi_{i+1,j}^{(0)})-y(y^{2}-3)\Xi_{i+1,j}^{(1)}-(y^{2}-3)^{2}\Xi_{i+1,j}^{(2)}
\end{equation}

Now we focus on the first $n-2$ rows. In particular, the row manipulation
already transforms the first column into the required form which can
be seen by explicit calculation (only the upper two elements are non-zero). Then we can complete the proof by induction after carrying
out the column operation, which has no effect on the first column.
The induction step is that the column operation up to the $j^{\mathrm{th}}$
column transforms the $(j-1)^{\mathrm{th}}$ column into the required
form, thus after the next addition, what we really add to the $j^{th}$
column is $y(y^{2}-3)(\Xi_{i,j-1}^{(0)}+y\Xi_{i,j-1}^{(1)})$. By
this we get
\begin{eqnarray}
\Xi_{i,j}^{(0)}+y\Xi_{i,j}^{(1)}+(y^{2}-3)\left[\Xi_{i,j}^{(2)}-\Xi_{i+1,j}^{(0)}+3\Xi_{i,j-1}^{(1)}\right]-\qquad\qquad\nonumber \\
-y(y^{2}-3)\left[\Xi_{i+1,j}^{(1)}-\Xi_{i,j-1}^{(0)}\right]-(y^{2}-3)^{2}\left[\Xi_{i+1,j}^{(2)}-\Xi_{i,j-1}^{(1)}\right]
\end{eqnarray}
where it can be shown that all the $[\,]$ brackets vanish separately
by using their definitions and relabeling the summation indices. We
can conclude that the upper-left $\left(n-2\right)\times\left(n-2\right)$
submatrix what we got by applying the operations to $\Xi\left(y_{+},y_{-},y_{1},\dots,y_{n}\right)$
is nothing but $\Xi\left(y,y_{1},\dots,y_{n}\right)=\Xi^{(0)}(n)+y\Xi^{(1)}(n)$,
and because the last column (except for its bottom element) only contains
terms proportional to $\sigma_{k}\left(y_{1},\dots,y_{n}\right)$
with $k>n$, they all vanish.

Now we have to prove that the bottom right element of the modified
matrix is $D_{n}$ which is done by induction. Have we applied only
the column operations the $(n-2,n-2)$ element would be by induction
$D_{n-1}$. Denote the $(n-1,n-2)$ element which is below this one
by $A_{n}$ what is a cumulative sum (of the last row's elements)
made by the column operations. This element is not effected by the
row operations. Now we apply the row operations: we add $\left(3-y^{2}\right)A_{n}$
to $D_{n-1}$, and - as explained in the previous paragraph - this
must produce
\begin{equation}
D_{n-1}\left(n\right)+\left(3-y^{2}\right)A_{n}=\Xi_{n-2,n-2}^{(0)}(n)+y\Xi_{n-2,n-2}^{(1)}(n)\label{eq:A4}
\end{equation}
On the other hand the only effect of the algorithm on the original
bottom right element $\Xi_{n-1,n-1}^{(0)}(n)+y\Xi_{n-1,n-1}^{(1)}(n)+(y^{2}-3)\Xi_{n-1,n-1}^{(2)}(n)$
is the addition of $y\left(y^{2}-3\right)A_{n}$ which should turn
it into something that we conjecture to be $D_{n}$. Now substituting
the form of $A_{n}$ from (\ref{eq:A4}), after some algebra, we get
\begin{eqnarray}
\Xi_{n-1,n-1}^{(0)}+y\Xi_{n-1,n-1}^{(1)}+(y^{2}-3)\Xi_{n-1,n-1}^{(2)}-y\Xi_{n-2,n-2}^{(0)}-y^{2}\Xi_{n-2,n-2}^{(1)}=\qquad\nonumber \\
=D_{n}-yD_{n-1}\label{eq:DynFinal}
\end{eqnarray}
The left hand side of (\ref{eq:DynFinal}) is manifestly zero which
can be checked explicitly after shifting summation indices. As the
$D_{n}$ polynomials are also thought of as functions of the $\sigma_{k}$
symmetric polynomials (\ref{eq:Dsigma}) it is easy to see that the
right hand side also vanishes.

\subsection{Kinematical recursion}

The desired form into which $\Xi_{n+2}\left(y,-y,y_{1},\dots,y_{n}\right)$
should be reduced is
\begin{equation}
\left(\begin{array}{ccc|cc}
 &  &  & 0 & 0\\*
 & \Xi\left(y_{1},\dots,y_{n}\right) &  & \vdots & \vdots\\
 &  &  & 0 & 0\\
\hline * & \dots & * & K_{n} & L_{n}\\
* & \dots & * & M_{n} & N_{n}
\end{array}\right)
\end{equation}
where the $0$s denote an $\left(n-3\right)\times2$ zero matrix and
the right bottom $2\times2$ submatrix satisfies $K_{n}N_{n}-L_{n}M_{n}=P_{n}\left(y\vert y_{1},\dots,y_{n}\right)$
as after taking the determinant it would reduce to the kinematical
recursion equation.

In this case we also apply row and column operations. The column operation
is as follows: take the first and second column, the third and fourth,
etc. and form pairs out of them, then add the first pair times $y^{2}\left(y^{2}-3\right)^{2}$
to the second pair (the first to the third, the second to the fourth),
then the new second pair to the third pair, and so on (each time take
the latest modified pair, and add it to the next pair after multiplying
it by $y^{2}\left(y^{2}-3\right)^{2}$). The row operation consists
of adding to each row $y^{2}$ times the row below and $y^{2}\left(y^{2}-3\right)$
times the second row below of the original matrix. Unlike to the column
operation it is not a cumulative sum. The order of the operations
is again arbitrary.

First we expand the elementary symmetrical polynomials in $\Xi\left(y,-y,y_{1},\dots,y_{n}\right)$
according to (\ref{eq:A1}) 
\begin{equation}
\Xi\left(y,-y,y_{1},\dots,y_{n}\right)=\Xi\left(n\right)-y^{2}\Xi^{\left(2\right)}\left(n\right)
\end{equation}
and we again proceed the proof by induction similarly to the dynamical
case. Now we focus on the first $n-3$ rows of this matrix. We execute
the row operation which brings the first two columns into the required
form and since they are not affected by the column operation we can
start the induction on the columns. We cumulatively add $y^{2}\left(y^{2}-3\right)^{2}$
times the $\left(j-2\right)^{\mathrm{th}}$ column to the $j^{\mathrm{th}}$
one and as the induction assumption we suppose that the $\left(j-2\right)^{\mathrm{th}}$
column have had already the good form (i. e. the same as the $\left(j-2\right)^{\mathrm{th}}$
column of $\Xi\left(n\right)$). After the row operation $(i,j)$
element becomes
\begin{equation}
\Xi_{i,j}-y^{2}\Xi_{i,j}^{(2)}+y^{2}\Xi_{i+1,j}-y^{4}\Xi_{i+1,j}^{(2)}+y^{2}\left(y^{2}-3\right)\Xi_{i+2,j}-y^{4}\left(y^{2}-3\right)\Xi_{i+2,j}^{(2)}
\end{equation}
Now applying the column operation and using the induction assumption
we get finally for $1\leq i,j\leq n-3$
\begin{eqnarray}
\Xi_{i,j}+y^{2}\left[\Xi_{i+1,j}-\Xi_{i,j}^{(2)}\left(n\right)-3\Xi_{i+2,j}+9\Xi_{i,j-2}\right]+\qquad\qquad\qquad\qquad\nonumber \\
+y^{4}\left[\Xi_{i+2,j}-\Xi_{i+1,j}^{(2)}+3\Xi_{i+2,j}^{(2)}-6\Xi_{i,j-2}\right]+y^{6}\left[\Xi_{i,j-2}-\Xi_{i+2,j}^{(2)}\right]
\end{eqnarray}
again each $[\,]$ bracket vanishes separately which can be seen by
shifting the summation indices in the elements of $\Xi$. This proves
that after the reduction of the determinant of $\Xi\left(y,-y,y_{1},\dots,y_{n}\right)$
its upper-left $(n-3)\times(n-3)$ block is $\Xi\left(n\right)$.
This argument is valid for the whole $\Xi\left(y,-y,y_{1},\dots,y_{n}\right)$
matrix except for the two bottom rows as in this cases the row operation
is different. In the last two columns of this $\left(n-3\right)\times2$
submatrix we have only terms which are proportional to $\sigma_{n+1}\left(y_{1},\dots,y_{n}\right)$
or $\sigma_{n+2}\left(y_{1},\dots,y_{n}\right)$ which are zeros,
therefore we have an $\left(n-3\right)\times2$ zero-block at the
upper-right corner.

Now all that is left to prove that the determinant formed by the bottom
right $2\times2$ elements is equal to $P_{n}\left(y\vert y_{1},\dots,y_{n}\right)$.
For that reason consider the $\left(n-2\right)\times\left(n-2\right)$
matrix $\Xi\left(y,-y,y_{1},\dots,y_{n-1}\right)$ and apply the row
and column operations. As already proved we end up with a matrix of
the form

\begin{equation}
\left(\begin{array}{c|c}
\Xi\left(n-1\right) & \begin{array}{cc}
0\quad & \quad0\end{array}\\
\hline \begin{array}{cc}
* & *\\
* & R_{n-1}
\end{array} & \begin{array}{cc}
K_{n-1} & L_{n-1}\\
M_{n-1} & N_{n-1}
\end{array}
\end{array}\right)\label{eq:Mat1}
\end{equation}
Now take the $\left(n-1\right)\times\left(n-1\right)$ matrix $\Xi\left(y,-y,y_{1},\dots,y_{n}\right)$.
Have we applied all the operation except from the addition of the
last row to any other row would result in (\ref{eq:Mat1}) as the
upper left block. Now apply this last row operation which results
in

\begin{gather}
\left(\begin{array}{cc|cc}
\begin{array}{c}
\Xi\left(n-1\right)\end{array} & 0 & 0 & *\\
* & K_{n-1} & L_{n-1} & *\\
\hline * & M_{n-1} & N_{n-1} & *\\
* & R_{n} & M_{n} & N_{n}
\end{array}\right)\longrightarrow\left(\begin{array}{c|c}
\Xi\left(n\right) & \begin{array}{cc}
0\  & \ 0\end{array}\\
\hline \begin{array}{cc}
* & *\\
* & R_{n}
\end{array} & \begin{array}{cc}
K_{n} & L_{n}\\
M_{n} & N_{n}
\end{array}
\end{array}\right)
\end{gather}
The only difference between the matrices is that we added $y^{2}$
times the last row to the one above and also added $y^{2}\left(y^{2}-3\right)$
times the last row to the second to the last row. Then we got the
relations
\begin{eqnarray}
K_{n-1}+y^{2}\left(y^{2}-3\right)R_{n} & = & \Xi_{n-3,n-3}\nonumber \\
L_{n-1}+y^{2}\left(y^{2}-3\right)M_{n} & = & 0\nonumber \\
N_{n-1}+y^{2}M_{n} & = & K_{n}
\end{eqnarray}
Two other relations can be derived if we consider how we got the $L_{n}$
and $N_{n}$ elements: we took the original elements at the kinematical
pole, added the $y^{2}\left(y^{2}-3\right)^{2}$ times the $\left(n-3\right)^{\mathrm{th}}$
column and $y^{2}$ times the row below which leads to 
\begin{eqnarray}
N_{n} & = & \Xi_{n-1,n-1}-y^{2}\Xi_{n-1,n-1}^{\left(2\right)}+y^{2}\left(y^{2}-3\right)^{2}R_{n}\nonumber \\
L_{n} & = & -y^{2}\Xi_{n-2,n-1}^{\left(2\right)}+y^{2}N_{n}+y^{2}\left(y^{2}-3\right)^{2}M_{n-1}
\end{eqnarray}
After eliminating $R_{n}$ and using (\ref{eq:XiShifted}) we get
\begin{eqnarray}
N_{n}=\sigma_{n}-\left(y^{2}-3\right)K_{n-1} & \quad;\quad & M_{n}=-\frac{1}{y^{2}\left(y^{2}-3\right)}L_{n-1}\label{eq:KLMNrec}\\
K_{n}=N_{n-1}-\frac{1}{y^{2}-3}L_{n-1} & \quad;\quad & L_{n}=-y^{2}\left(y^{2}-3\right)K_{n-1}+y^{2}\left(y^{2}-3\right)^{2}M_{n-1}\nonumber 
\end{eqnarray}
It can be also rearranged to a fourth order recursion
\begin{equation}
L_{n}=-y^{2}\left(y^{2}-3\right)\sigma_{n-2}+\left(6-y^{2}\right)L_{n-2}-\left(y^{2}-3\right)^{2}L_{n-4}\label{eq:Lfourthorder}
\end{equation}
Let our induction hypothesis be
\[
P_{n-1}=\det\mathcal{M}_{n-1}=K_{n-1}N_{n-1}-L_{n-1}M_{n-1}
\]
which can be easily checked for small values of $n$. With the relations
(\ref{eq:KLMNrec}) we can reformulate our conjecture as
\[
P_{n}+\left(y^{2}-3\right)P_{n-1}=\sigma_{n}N_{n-1}-\frac{\sigma_{n}}{y^{2}-3}L_{n-1}
\]
If we take into account (\ref{eq:Psigma}) it reduces to
\[
\sum_{p=1}^{n}\mathcal{P}_{p,0}\sigma_{n-p}=N_{n-1}-\frac{1}{y^{2}-3}L_{n-1}=\sigma_{n-1}+3\left(N_{n-3}-\frac{1}{y^{2}-3}L_{n-3}\right)+L_{n-3}
\]
which is indeed true for small values of $n$. At the second equation
we used (\ref{eq:KLMNrec}). By induction suppose that we already
proved that
\[
\sum_{p=1}^{n-2}\mathcal{P}_{p,0}\sigma_{n-2-p}=N_{n-3}-\frac{1}{y^{2}-3}L_{n-3}
\]
and what remains is to prove that
\begin{equation}
L_{n}=\sum_{p=1}^{n+3}\mathcal{P}_{p,0}\sigma_{n+3-p}-3\sum_{p=1}^{n+1}\mathcal{P}_{p,0}\sigma_{n+1-p}-\sigma_{n+2}\label{eq:Lconj}
\end{equation}
A lengthy but straightforward calculation shows that the right hand
side of (\ref{eq:Lconj}) satisfies the recursion (\ref{eq:Lfourthorder})
and one can check explicitly that for small $n$ the two sides of
(\ref{eq:Lconj}) are indeed equal that proves the validity of (\ref{eq:Lconj}),
and so the conjecture.

\end{document}